\documentclass[twocolumn]{aastex62}

\usepackage{lineno}
\usepackage{xcolor, soul}
\sethlcolor{pink}
\soulregister\citep7
\soulregister\citet7
\soulregister\ref7

\usepackage{multirow}
\graphicspath{{./}{figures/}}

\defcitealias{2017ApJ...850...22L}{LS17}
\defcitealias{2023ApJ...942....9H}{H23}
\defcitealias{2023ApJ...955L...6Y}{Y23}

\shorttitle{MIRONG Host Galaxies}
\shortauthors{Dodd et al.}

\begin{document}

\title{Mid-Infrared Outbursts in Nearby Galaxies: Nuclear Obscuration and Connections to Hidden Tidal Disruption Events and Changing-Look Active Galactic Nuclei}

\author[0000-0002-3696-8035]{Sierra A. Dodd}
\affiliation{Department of Astronomy and Astrophysics,
University of California, 
Santa Cruz, CA,  95064, USA }

\author{Arya Nukala}
\affiliation{Castilleja School, Palo Alto, CA, 94301, USA}
\affiliation{Department of Astronomy and Astrophysics,
University of California,
Santa Cruz, CA,  95064, USA }

\author{Isabel Connor}
\affiliation{Department of Astronomy and Astrophysics, University of California,
Santa Cruz, CA,  95064, USA }

\author[0000-0002-4449-9152]{Katie Auchettl}
\affiliation{School of Physics, The University of Melbourne, Parkville, VIC 3010, Australia}
\affiliation{ARC Centre of Excellence for All Sky Astrophysics in 3 Dimensions (ASTRO 3D)}
\affiliation{Department of Astronomy and Astrophysics, University of California, Santa Cruz, CA,  95064, USA }

\author[0000-0002-4235-7337]{K.D. French}  
\affiliation{Department of Astronomy, University of Illinois, 1002 W. Green St., Urbana, IL, 61801, USA}  

\author[0000-0001-8825-4790]{Jamie A.P. Law-Smith}
\affiliation{Department of Astronomy and Astrophysics, University of Chicago, Chicago, IL, 60637, USA}
\affiliation{Kavli Institute for Cosmological Physics, University of Chicago, Chicago, IL, 60637, USA}

\author[0000-0002-5698-8703]{Erica Hammerstein}
\affiliation{Department of Astronomy, University of Maryland, College Park, MD 20742, USA}
\affiliation{Astrophysics Science Division, NASA Goddard Space Flight Center, 8800 Greenbelt Road, Greenbelt, MD 20771, USA}
\affiliation{Center for Research and Exploration in Space Science and Technology, NASA/GSFC, Greenbelt, MD 20771, USA}

\author[0000-0003-2558-3102]{Enrico Ramirez-Ruiz}
\affiliation{Department of Astronomy and Astrophysics, University of California, Santa Cruz, CA,  95064, USA }
 
\begin{abstract}
We study the properties of galaxies hosting mid-infrared outbursts in the context of a catalog of five hundred thousand galaxies from the Sloan Digital Sky Survey. We find that nuclear obscuration, as inferred by the surrounding dust mass, does not correlate with host galaxy type, stellar properties (e.g. total mass and mean age), or with the extinction of the host galaxy as estimated by the Balmer decrement. This implies that nuclear obscuration may not be able to explain any over-representation of tidal disruption events in particular host galaxies. We identify a region in the galaxy catalog parameter space that contains all unobscured tidal disruption events but only harbors $\lesssim $ 11\% of the mid-infrared outburst hosts. We find that mid-infrared outburst hosts appear more centrally concentrated and have higher galaxy S\'ersic indices than galaxies hosting active galactic nuclei (AGN) selected using the BPT classification. We thus conclude that the majority of mid-infrared outbursts are not hidden tidal disruption events but are instead consistent with being obscured AGN that are highly variable, such as changing-look AGN.
\end{abstract}


\section{Introduction} \label{sec:intro}
The luminosity of supermassive black holes residing in the nucleus of most if not all galaxies is directly related to the rate at which they are supplied with matter. The fraction of supermassive black holes (SMBHs) that are in a highly luminous state has been observed to peak a few billion years after the Big Bang \citep{2001AJ....121...54F,2014MNRAS.437.3550M} 
and to gradually decline to the present day, where only one black hole out of every one hundred radiates close to its maximum allowed luminosity \citep{2008ARA&A..46..475H}. 
However, most SMBHs in the local universe still show some degree of activity \citep{2014ARA&A..52..589H}, ranging from outbursts with modest luminosities to highly luminous flares.

Black hole activity in the local universe has previously been associated with gas-rich mergers \citep{2001ApJ...555..719C}, which is thought to power SMBH growth in the early universe. However, AGN activity has been shown not to correlate with merger activity 
\citep[demonstrated for type II Seyfert hosts in][]{2003MNRAS.346.1055K}, and many SMBHs live in gas-poor environments that may be incapable of powering  highly variable outbursts \citep{2008ARA&A..46..475H,2013ARA&A..51..511K}. 
This is one of the main reasons why tidal disruption events (where a star approaches a SMBH close enough to be torn apart by tidal forces) are commonly invoked to explain some highly variable SMBH activity in the local universe \citep{2006ApJ...652..120M,2018ApJ...852...37A}. It has been recently argued that SMBHs in the local universe accrete depending on how recent their last episode of star formation occurred \citep{2003MNRAS.346.1055K}, which also seems to have profound consequences for moderating the rate of tidal disruption events \citep{2017ApJ...835..176F,2017ApJ...850...22L,2020SSRv..216...32F,2021ApJ...907L..21D}.

At high redshift, black hole activity is primarily driven by the accretion of gas, which is plentiful in the early universe as compared to today \citep{2005SSRv..116..523F}. Thus, there clearly exists a transition in how black holes are fed when the gas content of galaxies is drastically reduced \citep{2012MNRAS.427.3103B,2012MNRAS.419...95M}. Yet, the material surrounding accreting supermassive black holes is thought to be related to the active galactic nucleus with its host galaxy \citep{2015ApJ...811...26T}. For this reason, to probe the AGN–host galaxy connection directly in the local Universe, one needs to understand the structure and kinematics of the parsec-scale dust and gas that surrounds accreting SMBHs \citep{2013ARA&A..51..511K}.

During the last decade, mid-infrared interferometry has represented a major step forward in the characterization of nuclear dust in nearby AGNs \citep{2017NatAs...1..679R}.
Our current understanding of the close environment of accreting supermassive black holes obtained from infrared and X-ray studies of local active galactic nuclei suggests that the structure of the surrounding gas is complex, clumpy, and highly variable \citep{2017NatAs...1..679R,2018ARA&A..56..625H,2021ApJ...912...91T}.

Although scarcely explored, mid-infrared outbursts hold the potential of revealing some of the most dramatic obscured AGN activity, such as the disruptions of stars by SMBHs. What is more, they can be used to directly probe the parsec-scale dust and gas content in nearby AGN and explore the AGN–host galaxy connection for highly variable accretion episodes in nearby galaxies. In this {\it Letter}, we study the properties of galaxies hosting mid-infrared outbursts in the local Universe, presented in \S \ref{subsec:mirong_description}, in the context of a catalog of five hundred thousand galaxies from the Sloan Digital Sky Survey, described in \S \ref{subsec:catalog}, with the goal of constraining their origin. 

\section{Methods} \label{sec:methods}
\subsection{Reference Catalog} \label{subsec:catalog}
We use the galaxy catalog from \citet{2017ApJ...850...22L} and \citet{2021ApJ...907L..21D}. It consists of $\approx 5 \times 10^5$ galaxies from the Sloan Digital Sky Survey Data Release 7 MPA-JHU catalog\footnote{\url{http://wwwmpa.mpa-garching.mpg.de/SDSS/DR7}} \citep{2004MNRAS.351.1151B} with additional derived properties from \citet{2011ApJS..196...11S} and \citet{2014ApJS..210....3M}, yielding a wide range of host galaxy properties. These include velocity dispersion, emission line fluxes, Lick H$\delta_A$, and star formation rate (SFR) from the MPA-JHU catalog; redshift, bulge $g-r$, bulge and galaxy magnitudes, galaxy half-light radius, galaxy S\'ersic index, bulge to total light fraction, galaxy asymmetry indicator and galaxy inclination from \citet{2011ApJS..196...11S}; and bulge and total stellar masses from \citet{2014ApJS..210....3M}. Line fluxes in the catalog are calculated using the methodology described in Section 2.1 of \citet{2004ApJ...613..898T} and are corrected for stellar absorption features. We derive SMBH masses using the $M_{\rm bh}$--$\sigma_{e}$ scaling relation from \citet{2013ARA&A..51..511K}. We refer the reader to \citet{2017ApJ...850...22L} for additional discussion. 

A sample of AGN is selected from the host galaxy catalog using the BPT classification \citep{2003MNRAS.346.1055K,2006MNRAS.372..961K}, which considers the relative strength of OIII (5007\AA), $\textrm H\beta$, NII (6584\AA), and $\textrm H\alpha$ emission lines to infer AGN activity. We require the signal-to-noise for each of these lines to be $\geq 3$, yielding a final sample of $\approx 5 \times 10^4$ AGN.

\subsection{MIRONG} 
\label{subsec:mirong_description}
\citet{2021ApJS..252...32J}  conducted a systematic search of low-redshift ($z<0.35$)  galaxies that sustained mid-infrared outbursts based on {\it Wide-field Infrared Survey Explorer} (WISE) light curves, yielding a sample of 137 mid-infrared outbursts in nearby galaxies (MIRONG). 103 of the 137 MIRONG from \citet{2021ApJS..252...32J} are contained in our reference catalog and constitute our MIRONG sample. They are listed in Table \ref{tab:info}. Because the \citet{2021ApJS..252...32J} MIRONG sample is constructed from galaxies in the SDSS spectroscopic catalog with $z < 0.35$, we recover a high fraction of matches with the original sample. 

Of our 103 MIRONG sample, 3 were previously reported turn-on changing-look AGN (CL AGN): J0915+4814, J1133+6701, and J1115+0544 \citep{2019ApJ...883...31F,2018ApJ...862..109Y,2019ApJ...874...44Y}. We also recover 4 unclassified optical transients (J0045-0047, J0841+0526, J1533+2729, J1647+3843) and 1 spectroscopically-confirmed supernova (J1540+0054/ ASASSN-16eh). 

\citet{2022ApJS..258...21W} obtain multi-epoch follow-up spectra of 54 of the 137 originally reported  MIRONG, of which 43 are in our host galaxy sample. They propose tentative classifications of either turn-on CL AGN, AGN flare, or TDE for the 22 MIRONG exhibiting emission line variability, 16 of which are in our sample. Based on their analysis, 9 of the 16 in our sample are tentatively classified as TDEs, 5 as non-specified AGN flares, and 2 as turn-on CL AGN. 

Studying the host galaxy properties of MIRONG provides us with an alternative classification scheme to constrain their possible identities, for those with and without spectroscopic follow-up. This can be effectively done if nuclear obscuration is fairly independent of galaxy type. We turn our attention to this critical issue in Section~\ref{sec:dust}. In summary, we set out to analyze the host galaxy properties of 103  MIRONG, 43 of which have additional follow-up spectroscopy as described in \citet{2022ApJS..258...21W}, with the goal of further understanding their hidden origin.  

\begin{figure*}
\plotone{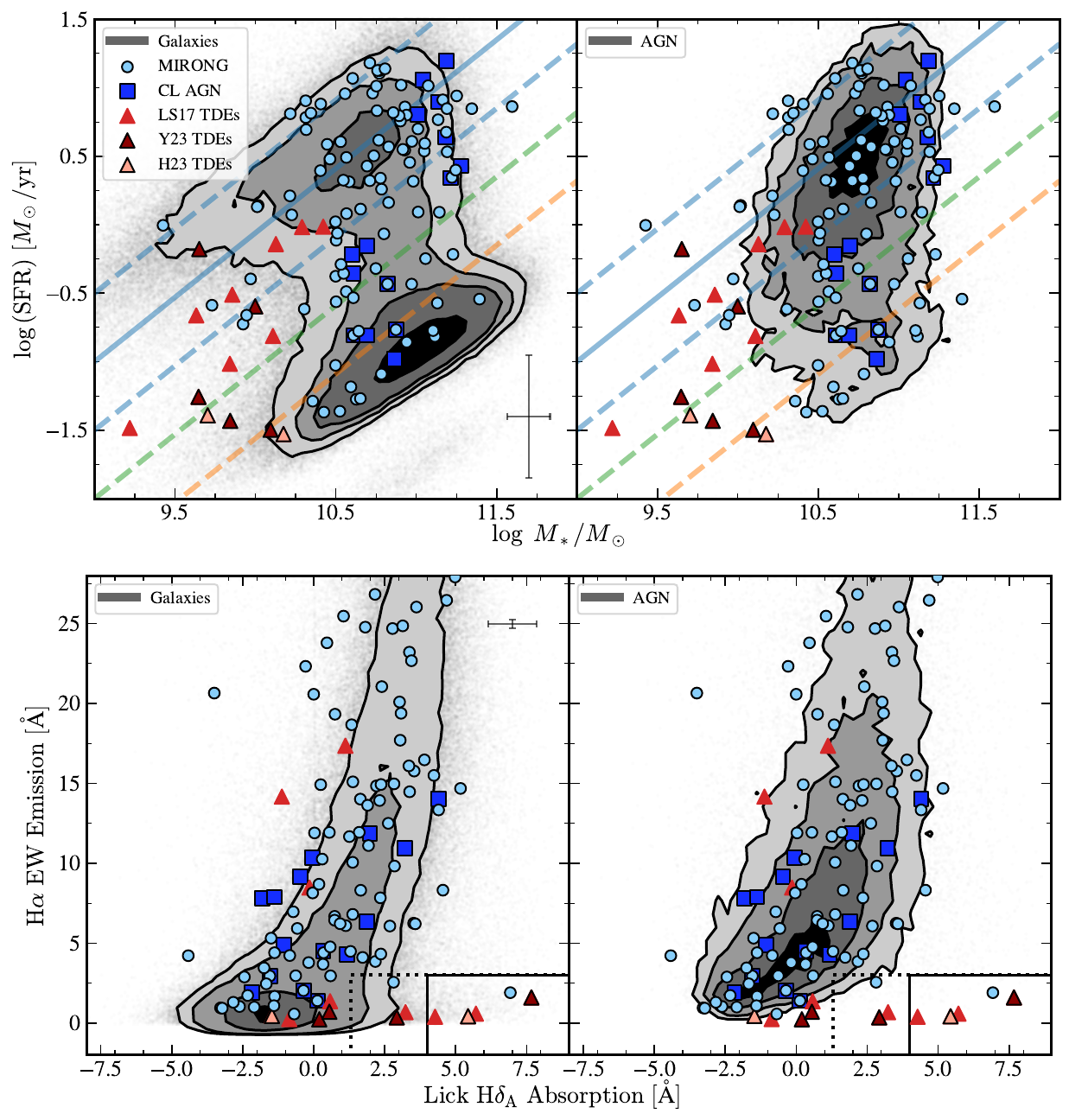}
\caption{\textbf{Top panels:} Star formation rate (SFR) vs. total stellar mass ($M_*$) with galaxies (left) and AGN (right) as contours, with MIRONG (light blue circles), CL AGN (dark blue squares), and TDEs (all triangles; \citet{2017ApJ...850...22L} in red, \citet{2023ApJ...942....9H} in pink, and  \citet{2023ApJ...955L...6Y} in maroon). Following \citet{2017ApJ...850...22L} and \citet{2021ApJ...907L..21D}, the star-forming main sequence region is defined according to \citet{2010ApJ...721..193P} and designated by the solid blue line. Each line shown is separated by 1$\sigma$ in SFR. Descending in SFR, the green valley is denoted as the region between the lower blue dashed line and above the dashed orange line. The median error in SFR and $M_*$ is shown in the bottom right of the left plot. MIRONG appear to closely follow the general AGN population, with a particular grouping in the highly-star-forming region. 3  of the 4 listed turn-on MIRONG are contained in our CL AGN sample: J1115+0544, J0915+4814, and J1133+6701. \textbf{Bottom panels:} H$\alpha$ equivalent width (EW) versus Lick H$\delta_A$ absorption for the same populations. The rectangle created by the solid lines demarcates E+A galaxies, while the larger rectangle created by the dotted region more broadly designates quiescent Balmer-strong galaxies \citep{2016ApJ...818L..21F, 2017ApJ...850...22L}. Median error is shown in the top right of the left plot. For consistency with \citet{2017ApJ...850...22L}, the maximum H$\alpha$ EW is capped at 28 angstroms. However, 13 MIRONG have H$\alpha$ EW emission values exceeding this limit and thus are not shown. Such elevated values indicate that a subgroup of MIRONG is more highly-SF than typical AGN. Interestingly, 1 TDE from \citet{2023ApJ...955L...6Y} (AT2020neh) also has a H$\alpha$ EW value in excess 28 angstroms. We caution that a direct comparison between the contours shown and the galaxies hosting various transient populations can only be made when the underlying selection mechanisms are taken into account. This is particularly relevant for  TDE hosts (Section~\ref{sec:tde_comp}) but less important for CL AGN  and MIRONG hosts (Section~\ref{sec:appendix_a}). 
\label{fig:fig1}}
\end{figure*}

\section{Host Galaxy Properties} \label{sec:hosts}
We begin by considering the star formation rate (SFR) and total stellar mass, $M_\ast$, of the host galaxies of MIRONG in relation to our galaxy catalog (see the top panels of Figure \ref{fig:fig1}). We also include in our comparison a sample of 15 CL AGN from \citet{2021ApJ...907L..21D}, as well as a sample of TDEs. This includes 8 TDEs from \citet{2017ApJ...850...22L} and 7 TDEs from the  Zwicky Transient Facility (ZTF) that are contained in our reference catalog (AT2018hyz,  AT2019azh, AT2020ocn, AT2020wey, AT2020vwl, AT2021nwa, and AT2020neh; see, e.g., \citet{2023ApJ...942....9H,2023ApJ...955L...6Y}, and references therein). From here on out, we refer to these TDE sub-samples by the first initial and year of publication (i.e. \citetalias{2017ApJ...850...22L}, \citetalias{2023ApJ...942....9H}, and \citetalias{2023ApJ...955L...6Y}). We note that although there are TDEs in common between the works of \citetalias{2023ApJ...942....9H} and \citetalias{2023ApJ...955L...6Y}, we refer to any shared TDEs by \citetalias{2023ApJ...955L...6Y} for consistency with later host galaxy analysis of this sample in Section \ref{sec:tde_comp}. 

We define the star-forming main sequence (SFMS) of galaxies in this plane to fall along the solid blue line \citep{2010ApJ...721..193P}. Dashed lines spaced by 1$\sigma$ (the median scatter in the SFR measurements) are added to indicate degrees of quiescence. Contours of galaxies and AGN plotted in Figure \ref{fig:fig1} and throughout this analysis are spaced by 0.5$\sigma$ (11.8\%; note that percentages associated with $\sigma$ in 2D histograms differ from 1D).  

Two distinct groupings of galaxies are observed in the SFR - $M_\ast$ plane for galaxies (upper left panel of Figure \ref{fig:fig1}). The first, located in the top left and along the SFMS, consists of spiral, late-type galaxies. The SFMS provides a critical tool for studying the quenching of star formation \citep[e.g.,][]{2004MNRAS.351.1151B} and the possible emergence of quiescent galaxies (as indicated by the second grouping in the lower right). Although the exact process by which this transformation takes place is debated, galaxies falling between late- and early-type galaxies are understood to be in a state of transition \citep{2007ApJS..173..342M}, and are often referred to as \textit{green valley} galaxies. Following \citet{2017ApJ...850...22L} and \citet{2021ApJ...907L..21D}, we designate the green valley region as falling between the lower blue dashed line and the orange dashed line in the SFR - $M_\ast$ plane. 

Before we consider how the various populations relate to one another in this plane, it is important to consider any selection biases that could prevent fair comparison. MIRONG are selected with SDSS as the parent sample, which makes them consistent with our main catalog and AGN subset (see Figure \ref{fig:fig6} and Appendix \ref{sec:appendix_a} for a comparison of host stellar masses and redshift distributions of these samples). The selection methodology for TDEs in transient surveys, however, biases them toward fainter host galaxies than a typical galaxy in our catalog. For that reason, although we include the TDEs for context and consistency with previous works, we caution against directly comparing TDEs to MIRONG in this plane. We investigate the relationship between MIRONG and TDEs by constructing an unbiased galaxy catalog in Section \ref{sec:tde_comp}.

AGN are generally seen to activate preferentially in SF galaxies \citep{2008ARA&A..46..475H,2013ARA&A..51..511K}, and persist through (and possibly drive) the eventual quenching phase (see the upper right panel of Figure \ref{fig:fig1}). As expected, the distribution of MIRONG appears to be more closely aligned in the SFR - $M_\ast$ plane with AGN than galaxies. The MIRONG hosts also do not exhibit any large degrees of clumping or grouping like that seen in CL AGN hosts (although the larger sample size is certainly relevant), but the populations do appear to slightly overlap.

As clearly seen in Figure \ref{fig:fig1}, TDEs are observed to preferentially take place in post-starburst host galaxies, also known as E+A galaxies \citep{2016ApJ...818L..21F, 2017ApJ...850...22L}. These unique hosts can be identified through the H$\alpha$ equivalent width vs. Lick H$\delta_A$ absorption plane. H$\alpha$ equivalent widths are associated with current star formation, while Lick H$\delta_A$ absorption results primarily from A-type stars. E+A galaxies lie in the bottom right of this plane, as seen in the lower panels of Figure \ref{fig:fig1}, where low values of H$\alpha$ equivalent width indicate little to no ongoing star formation, and high values of Lick H$\delta_A$ absorption indicate a starburst in the last $\approx$ 1 Gyr. As in \citet{2016ApJ...818L..21F} and \citet{2017ApJ...850...22L}, we define E+A galaxies as residing in the rectangle created by the solid black lines in the lower panels of Figure \ref{fig:fig1} and Balmer-strong quiescent galaxies as those falling in the larger rectangle created by the dashed black lines. 

\citet{2017ApJ...850...22L} demonstrated that the overrepresentation of TDEs in post-starburst galaxies persists even after controlling for black hole mass, redshift, presence of a strong AGN, bulge colors, and surface brightness of host galaxies. \citet{2016ApJ...818L..21F} and \citet{2018ApJ...853...39G} also found rate enhancements of TDEs in quiescent Balmer-strong galaxies, the rates of which were demonstrated by \citet{2020SSRv..216...32F} to all be consistent with one another, although \citet{2021ApJ...908L..20H} found that additionally controlling for green valley preference can account for this overrepresentation in ZTF-I TDE hosts. Again, although we caution against drawing direct comparisons between the two samples given the TDE host preference for fainter galaxies (the reader is referred to Section \ref{sec:tde_comp}), MIRONG generally do not appear to share TDE hosts' over-representation in the quiescent Balmer-strong region. As in the SFR - $M_\ast$ plane, MIRONG hosts seem to broadly follow the distribution of AGN. 

Whether the over-representation of TDEs in post-starburst galaxies is physically driven or the result of observational biases is the subject of ongoing study. \citet{2017NatAs...1E..61T} suggest that the intrinsic rate of TDEs in highly SF galaxies can dominate over that in post-starburst galaxies, but that high nuclear dust content in such galaxies could make detection extremely difficult. \citet{2021ApJ...910...93R} also find that surveys such as ZTF may account for only $\sim$30\% of TDEs that would be detectable if no dust were present in their host galaxies. Are TDEs in SF galaxies preferentially obscured?  It is to this question that we next turn our attention to by studying nuclear obscuration as a function of galaxy type and interstellar extinction.   
\begin{figure*}
\epsscale{1.1}
\plotone{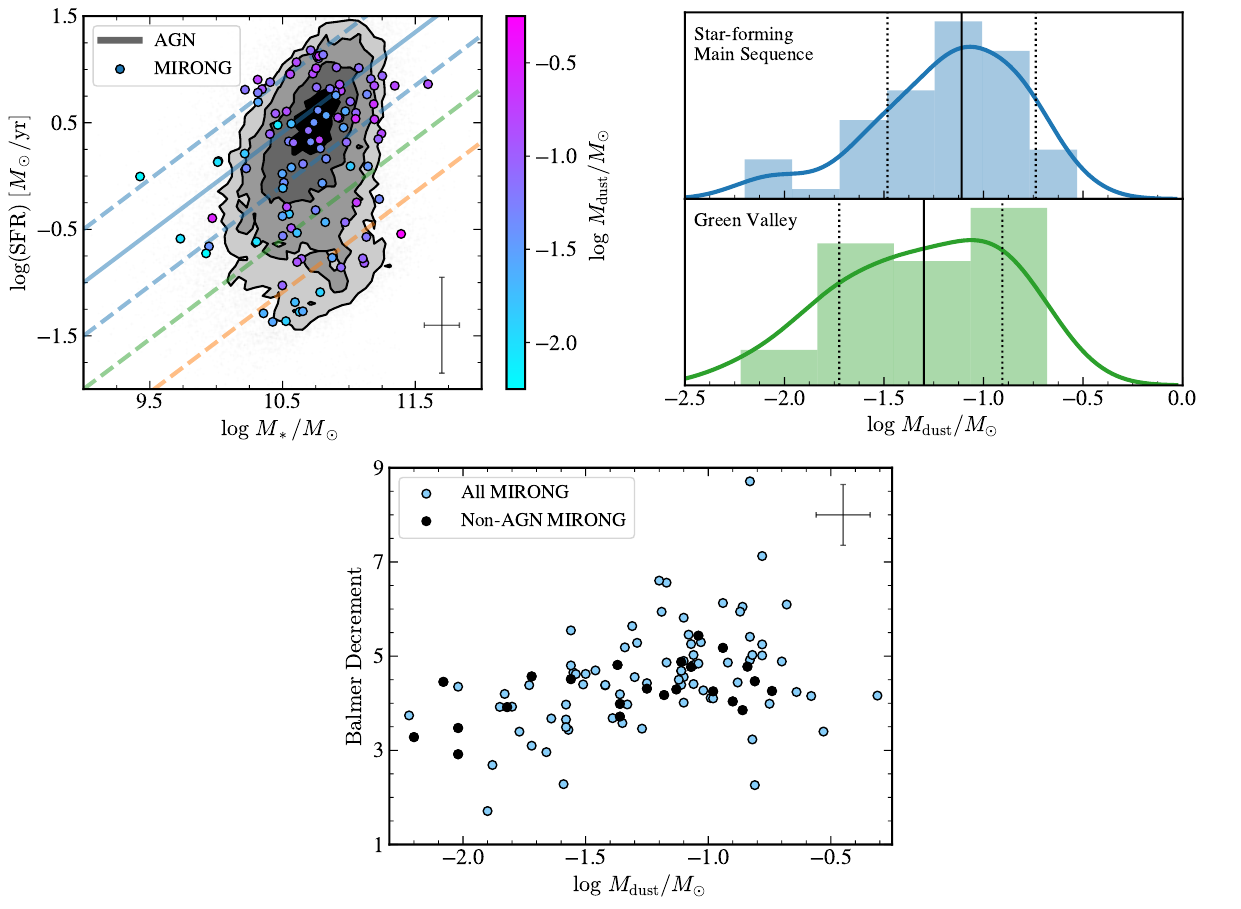}
\caption{\textbf{Top left:} Star formation rate vs. total stellar mass as seen in the top panel of Figure \ref{fig:fig1}, this time with only AGN contours and MIRONG scatter points. The colorbar indicates derived log dust mass from \citet{2021ApJS..252...32J} in units of $M_\odot$.
\textbf{Top right:} Histogram of log dust mass for MIRONG lying in the SFMS (top) and green valley (bottom). Solid lines represent the median of each distribution, and dotted lines are spaced $\pm$ 1$\sigma$ from the median. The two distributions are broadly consistent with one another, indicating that nuclear dust of MIRONG host galaxies does not vary with galaxy type. One outlying MIRONG (J154029.29+005437.2) has been excluded from all plots in this figure and is found in the green valley with a log dust mass of -3.71. Including this MIRONG only serves to broaden the standard deviations and leaves the median in place.
\textbf{Bottom:} Balmer decrement vs. log dust mass for MIRONG host galaxies. The error bar in the top right represents the median error value for all MIRONG host galaxies in each parameter. Errors for Balmer decrement measurements are calculated from the error in H$\alpha$ and H$\beta$ line flux measurements multiplied by the recommended MPA-JHU uncertainty scaling factors of 2.473 (for H$\alpha$) and 1.882 (for H$\beta$). 
\label{fig:fig2}}
\end{figure*}

\section{Nuclear Obscuration, Galaxy Type  and Galactic-Scale Extinction}\label{sec:dust}
The derived dust properties in the nuclear region of MIRONG hosts \citep{2021ApJS..252...32J} allow us to investigate the possibility that TDEs in high dust-content galaxies might be preferentially hidden. Dust mass is derived by assuming that thermal emission from dust is powering the MIR emission. The assumed dust grain size distribution comes from Mathis, Rumpl, and Nord \citep[MRN; ][]{1977ApJ...217..425M} and is roughly power-law in nature. W1 and W2 fluxes are then fit to a modified blackbody to obtain dust temperature and corresponding masses under the assumption of a high dust covering fraction \citep{2021ApJS..252...32J}. This assumption seems to be additionally supported by our findings that dust mass is not correlated with galaxy orientation in our galaxy host sample. Because the methodology for deriving dust mass estimates is applied uniformly to all MIRONG, additional factors such as possible differences in individual galaxy dust composition are not taken into account. As such we focus our following analysis on dust mass properties of MIRONG on a population rather than individual level.

To investigate how nuclear dust mass varies with galaxy type, we revisit the SFR - $M_\ast$ plane, now with each MIRONG shaded according to their dust mass content (in units of $\log \ M_\odot$; see the upper panels of Figure \ref{fig:fig2}). Contours of AGN are shown in the background. No obvious trend is visible between the location on the evolutionary sequence of galaxies and nuclear dust mass. As discussed in the previous section, it has been suggested that TDEs could occur at elevated rates in highly SF galaxies but would be obscured by higher levels of nuclear dust 
\citep{2017NatAs...1E..61T}. Since TDEs are preferentially observed in green valley host galaxies, we compare the nuclear dust content of SFMS and green valley MIRONG galaxies in the upper right panel of Figure \ref{fig:fig2} to see whether nuclear dust levels are elevated in highly SF galaxies. The two distributions are similar, as also seen by their median values (solid black lines) and $\pm 1\sigma$ (dashed black lines). We perform a Kolmogorov-Smirnov test on the two samples. The resultant p-value of 0.37 implies that we are unable to reject the null hypothesis that the two populations are drawn from the same underlying distribution, suggesting they are consistent with one another. This is in stark contrast to predictions that advocate for an elevated rate of obscured TDEs in SF galaxies. We also find no trend between $\rm D_n(4000)$, an indicator of the age of the galaxy stellar population \citep{2003MNRAS.341...33K}, and nuclear dust mass. We can thus conclude that nuclear obscuration does not correlate with stellar properties such as stellar mass and age. Combined with the finding that nuclear dust does not depend on SFR, we can conclude that obscured TDEs would not prefer one type of galaxy over another. This implies we are not missing a population of TDEs in more highly-SF galaxies, and that their observed green valley preference might be based on underlying physical mechanisms controlling the rate of TDEs \citep[see, e.g.,][and references therein]{ 2017ApJ...850...22L}. 
 
Having shown that nuclear obscuration does not depend on galaxy type, here we consider how the Balmer decrement, which is thought to correlate with the dust mass content in a galaxy, correlates with the abundance of dust surrounding SMBHs. Balmer decrement, which consists of the ratio between H$\alpha$ and H$\beta$ emission lines from the Balmer series (n=2), is commonly used as a measure of galaxy interstellar extinction. However, the presence of an AGN can alter the region probed by this measurement. To this end,  the Balmer decrement has been used to measure dust extinction in the broad line regions of type I AGN and quasars \citep{2008MNRAS.383..581D,2023MNRAS.522.5680M}, as well as in the narrow line region of AGN \citep{2019MNRAS.483.1722L}. For partially obscured AGN, such as the ones in our MIRONG sample, we can expect the Balmer decrement to reflect extinction on both nuclear and galactic scales. Because most of the MIRONG in our sample (76; 52 of which have strong ($\geq 3$) signal-to-noise ratio in the BPT line measurements) are classified as AGN in the BPT diagram, we expect a large degree of scatter in their Balmer decrements. 

The bottom panel of Figure \ref{fig:fig2} shows the Balmer decrement of all MIRONG vs. nuclear log dust mass, separated into galaxy hosts with and without AGN activity. As anticipated, we see a large degree of scatter in the AGN group. The non-AGN show considerably less scatter. Neither group shows a strong correlation between Balmer decrement and nuclear dust mass\footnote{This statement is supported by correlation coefficients $\lesssim 0.5$ for each population in both Pearson and Spearman tests (Correlation coefficient values: AGN MIRONG Pearson = 0.39, AGN MIRONG Spearman = 0.41, non-AGN MIRONG Pearson = 0.53, non-AGN MIRONG Spearman = 0.37.). We note that the highest correlation coefficient of 0.53 for non-AGN MIRONG could be consistent with a weak correlation with significant scatter.}. As the Balmer decrements of the non-AGN group are more likely to be indicative of galactic-scale dust, this suggests no clear correlation between nuclear- and galactic-scale dust in these systems. This is in agreement with works such as \citet{2014MNRAS.437.3550M}, which find that the degree of nuclear obscuration appears uncorrelated with larger-scale galactic properties including SFR and total stellar mass.  This provides an additional clue that the dust content of galaxies in the local Universe at the host scale does not determine their innermost dust content.

\section{On the Origin  of MIRONG sources }\label{sec:id}
Finding any correlation between nuclear- and galactic-scale dust content is key to understanding any potential host galaxy preference for the MIRONG population. With this information, we can examine the notion that MIRONG in highly SF regions (where galactic dust is expected to be abundant) could plausibly be obscured TDEs that until now we had no means of uncovering.
Having proven that nuclear obscuration does not depend on host galaxy type, it is clear that the hidden population should mirror the unobscured population. This enables us to use host galaxy properties of unobscured nuclear transients to identify possible MIRONG origins. In this Section, we explore possible explanations for the origin of the bulk of the MIRONG population, including the fraction of MIRONG that might be TDEs.

\subsection{CL AGN Matching in the SFR - $M_\ast$  Plane}
We identify MIRONG hosts located within a range of SFR and total stellar mass of CL AGN hosts using the matching methodology of \citet{2017ApJ...850...22L} and \citet{2021ApJ...907L..21D}, but with a 10\% tolerance window (corresponding to $0.21$ in $\log $ SFR and $0.07$ in $\log \ M_\ast$). As is customary, for this matching procedure and others that follow, tolerances are chosen to be consistent with the mean separation of sources in the plane of comparison. Table \ref{tab:tab1} summarizes our findings. We locate 18 MIRONG with matching properties to CL AGN hosts (including recovering the 3 duplicates), corresponding to 17.5\% of MIRONG in our sample.

Of the 18 MIRONG with similar host properties to CL AGN, 10 are classified as AGN in the catalog (19.2\%; see Section \ref{sec:methods} for details on AGN classification). For comparison, the relative percentages of AGN and galaxies near CL AGN hosts are 18.0\% and 14.7\%, respectively. This implies that the MIRONG population is about as likely to resemble a CL AGN host as a typical AGN. This is perhaps unexpected, given that MIRONG are selected such that they need to be associated with a highly variable nuclear source. Even though MIRONG do not stand out from AGN in this regard, we consider host galaxy S\'ersic index as a means of distinguishing between typical AGN and highly variable (HV) AGN (including CL AGN) in the next section. 

Most MIRONG hosts broadly resemble the general AGN population in the SFR and $M_\ast$ plane. We confirm this using a 2D Kolmogorov–Smirnov test and are thus unable to reject the null hypothesis that the two samples were drawn from the same distribution (p value: 0.11). This is however not the case when compared to the galaxy population (p value: $6\times 10^{-8}$). Despite their apparent similarity to typical AGN, the identification of MIRONG via their extreme outbursts points to a more dramatic type of variability than what is expected for typical  AGN in the local Universe. CL AGN represent the extreme end of these highly-variable AGN. A useful phenomenological distinction between AGN and highly variable (HV) AGN is galaxy S\'ersic index \citep{2021ApJ...907L..21D}. In the following section we thus analyze MIRONG in this context.

\subsection{S\'ersic Index Comparison}
S\'ersic index measures the light concentration of the galaxy surface brightness. Higher values correspond to highly centrally concentrated light profiles, such as those often exhibited by elliptical galaxies, while lower values are more consistent with diffuse profiles, such as those seen in spirals. S\'ersic index is generally thought to provide a measurement of the density profile of stars and, to a lesser extent, the kinematic state of the star-forming gas in the nuclear region \citep{2021ApJ...908..123R}. As mentioned previously, highly variable AGN and CL AGN have been shown to have higher S\'ersic indices than AGN \citep{2021ApJ...907L..21D}, even when controlling for black hole mass. TDE hosts also exhibit this trend to a slightly lesser degree \citep{2017ApJ...850...22L}. 

\begin{figure}
\plotone{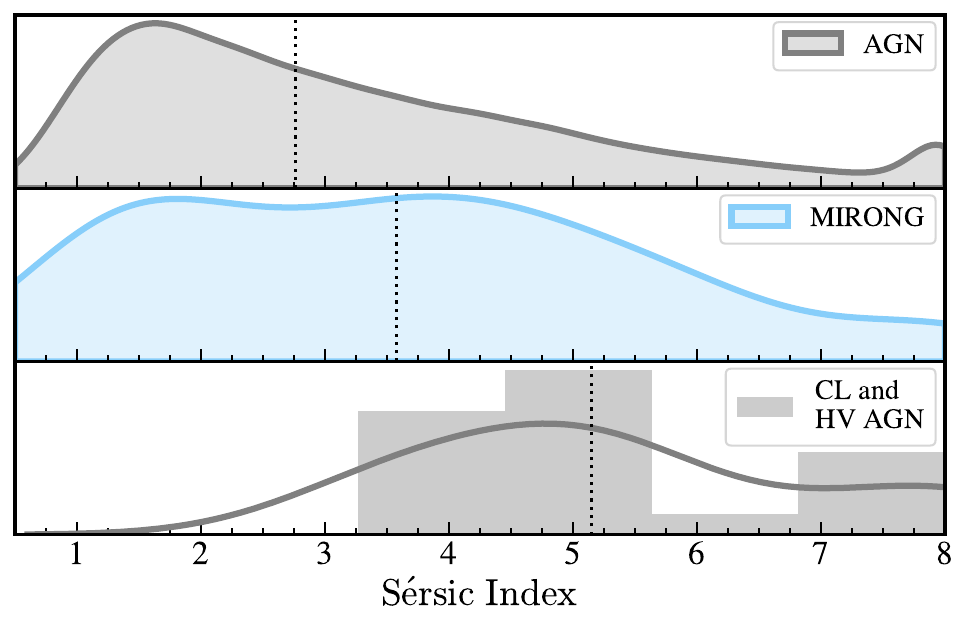}
\caption{ S\'ersic index smoothed histograms for AGN, MIRONG, and CL + HV AGN host galaxies. The area under the distributions is normalized to 1. Dashed vertical lines represent the median of each distribution. The MIRONG population generally display higher S\'ersic indices than typical AGN.  
\label{fig:fig3}}
\end{figure}

Figure \ref{fig:fig3} shows S\'ersic index for AGN, MIRONG, and CL and HV AGN. The four HV AGN used in our sample are the same from \citet{2021ApJ...907L..21D} and consist of KUG 1624+351 \citep{2014MNRAS.444.1041K}, J094608+351222 
\citep{2017MNRAS.470.4112G}, 2MASS J09392289+3709438 \citep{2016A&A...592A..74S}, and Swift J1200.8+0650 
\citep{2007ApJ...669..109L}. Intriguingly, the last object listed (Swift J1200.8+0650) is also a MIRONG source, even though it was selected independently. The MIRONG distribution differs markedly from that of AGN. MIRONG S\'ersic indices tend to fall between those of AGN and CL and HV AGN. We also performed this analysis with a sample of AGN matched to MIRONG based on SFR, total stellar mass, and redshift, and found no noticeable difference in the resulting distribution compared to typical AGN. We also note that although the distribution of S\'ersic indices for AGN is slightly higher for active versus inactive galaxies, this has been shown to become less notable for lower-resolution galaxies (with redshifts larger than $z > 0.05$) and thus likely does not strongly influence our findings here \citep{2008MNRAS.384..420G}. Comparing the distribution of MIRONG S\'ersic indices to these other populations suggests that HV AGN are the likeliest source of MIRONG behavior, as opposed to standard AGN-type flares.

\subsection{The rate of TDEs in the MIRONG population}\label{sec:tde_comp}

\begin{figure*}
\plotone{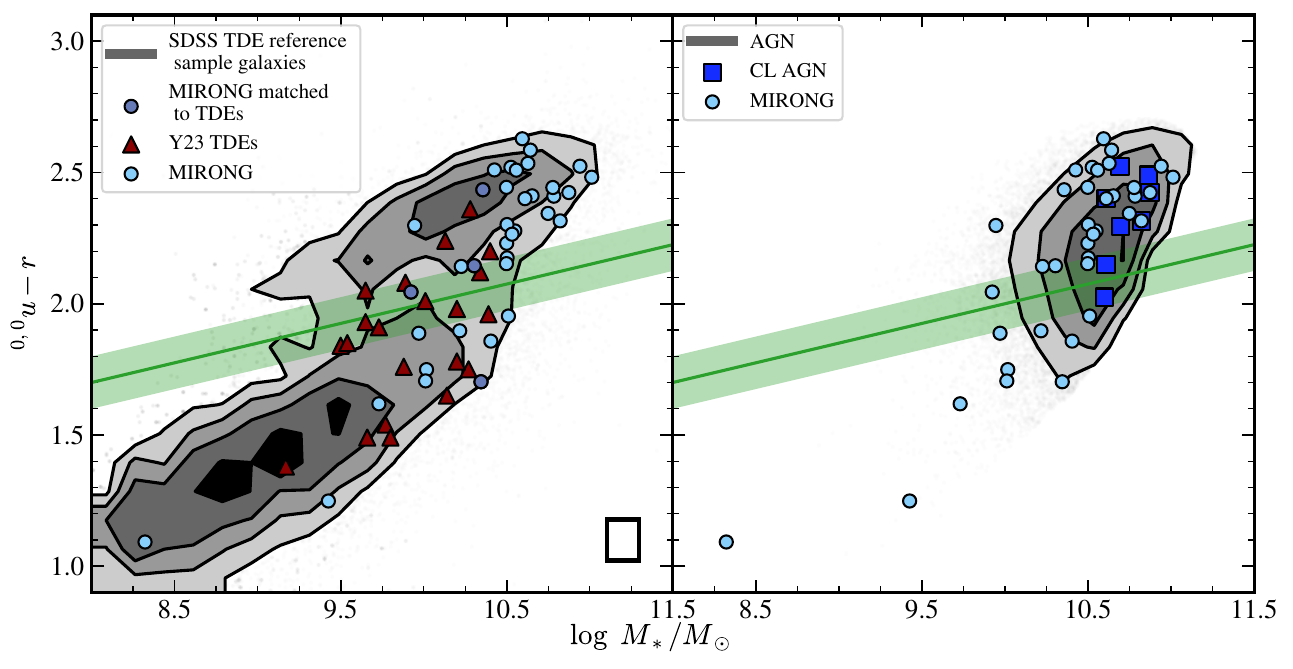}
\caption{Host galaxy color ($^{0,0}u-r$) versus total stellar mass for SDSS TDE reference samples (see Section \ref{sec:tde_comp} for a detailed explanation of sample selection) with the green valley region in this plane as defined by \citet{2023ApJ...955L...6Y}. \textbf{Left:} Galaxy contours with MIRONG, MIRONG near TDE hosts, and TDEs. Matching is performed using the methodology of \citet{2017ApJ...850...22L,2021ApJ...907L..21D} with a 9\% tolerance, yielding 4 MIRONG matches ($\lesssim 11\%$). A box representing the matching size is shown in the lower right. \textbf{Right:} AGN contours with MIRONG and CL AGN. We note that even with these additional, restrictive selection criteria, MIRONG generally follow the AGN sample. 
\label{fig:fig4}}
\end{figure*}

In order to compare TDE hosts to the MIRONG population, a control galaxy sample must be constructed. We employ the methodology described in Section 2.2 of \citet{2021ApJ...908L..20H} to create a sample of SDSS galaxies that can be directly compared to our TDE hosts. We derive the maximum redshifts that SDSS would be complete out to for each of the 33 \citetalias{2023ApJ...955L...6Y} TDEs\footnote{We note that we perform our analysis using \textit{g}- instead of \textit{r}-band values as done in \citet{2021ApJ...908L..20H}. Properties of the resulting galaxy sample are consistent regardless of band choice.}. As in \citet{2021ApJ...908L..20H}, we cap our SDSS galaxy matches per TDE to be 1000 and resample as needed for TDEs with less than 1000 matches. This constitutes our SDSS TDE reference sample and can be seen in the contours of the left panel of Figure \ref{fig:fig4}, which shows host galaxy color ($^{0,0}u-r$) versus total stellar mass, as in Figure 18 of \citet{2023ApJ...955L...6Y}. $^{0,0}u-r$ is defined as the difference between the rest frame, galactic extinction-corrected $u$ and $r$ absolute magnitudes from the \texttt{Photoz} table in SDSS DR7. The $^{0,0}u-r$ vs total stellar mass plane also allows for the identification of the green valley region, as described earlier in Section \ref{sec:hosts}, and also allows us to use the full sample of 33 \citetalias{2023ApJ...955L...6Y} (as opposed to only using the 5 contained within our original catalog that have SFR measurements). Here the green valley is defined according to Equation 22 from \citet{2023ApJ...955L...6Y}. TDEs are over-represented in the green valley of this plane \citep{2023ApJ...955L...6Y} as well as in the SFR versus stellar mass plane of Figure \ref{fig:fig1}. 

In order to robustly compare the host properties of TDEs in relation to the other populations, we perform the following steps. First, we select TDEs that fall at or within the outer contour of host galaxy color and total stellar mass shown in Figure~\ref{fig:fig4}. The outer contour contains 86.4\% of the SDSS TDE reference sample. Second, we select MIRONG, CL AGN, and AGN that fall within the same boundary. 
This ensures that the magnitude limit corrections described above are applied to each of these distinct samples. This yields 37/103 MIRONG (35.5\%), 21/33 TDEs (63.6\%), 8/15 CL AGN (53.3\%), and 29,375/52,613 AGN (55.8\%). We now perform the tolerance matching described in the previous section but for TDEs and MIRONG. We use a 9\% tolerance range (corresponding to $0.08$ in $^{0,0}u-r$ and $0.10$ in $\log \ M_\ast$). A sample box size can be seen in the lower right of the left panel of Figure \ref{fig:fig4}. We find 4 matches out of the 37 MIRONG in this SDSS TDE comparison sample (10.8\%). A comparison of total stellar masses of the SDSS TDE reference samples can be seen in Figure \ref{fig:fig7} and is explored in Appendix \ref{sec:appendix_b}. 

We can also consider the distribution of the SDSS TDE reference sample-limited AGN, CL AGN, and MIRONG in the host galaxy color versus total stellar mass plane. As seen on the right panel of Figure \ref{fig:fig4}, MIRONG again mirror the general AGN population and occupy a similar region to CL AGN. We note that even in this more restrictive sample that is limited in redshift to account for TDE selection biases, we still find that MIRONG hosts appear much more closely related to AGN than TDEs.

We now examine if our derived rate of TDEs in MIRONG based on host galaxy matching of $\lesssim$ 11\% is consistent with the classification of MIRONG from \citet{2022ApJS..258...21W}. The authors spectroscopically monitored 53 of the 103 original MIRONG sources over a roughly 4-year period. Of these 53, 22 (41.5\%) displayed variability in the broad H$\alpha$ emission line (EL). \citet{2022ApJS..258...21W} perform their subsequent classification on this subsample of 22 objects, based on the notion that any light from a transient event located at or near the central, obscured SMBH would have been reprocessed by the dust over this monitoring timescale.  

TDEs are selected from this group of 22 EL-variable sources as follows. Sources are split into likely AGN and quiescent groups. Quiescent sources are classified as TDEs with the exception of one turn-on AGN candidate. Any AGN MIRONG with iron coronal lines and He II$\lambda4686$ features are also tentatively classified as TDEs given the association of those features with TDE spectra \citep{2019ApJ...887..218L}. This results in 14 TDE candidates ($\approx 26\%$).

The authors highlight the challenges associated with classifying obscured nuclear transients. This includes the possibility of a myriad of origins for AGN flares, for which we have very few expected spectral templates for comparison (although much progress has been made recently, see, e.g., \citet{2023arXiv230703182C}). As such, we hope to offer a complementary view to the work of \citet{2022ApJS..258...21W} based on host galaxy properties of unobscured TDEs. We have 6 of the 14 \citet{2022ApJS..258...21W} TDE candidates in our entire galaxy catalog, but none within the SDSS TDE reference sample. Although we are unable to directly compare the TDE candidates from \citet{2022ApJS..258...21W} to our SDSS TDE reference sample, the overall relatively low fraction of MIRONG with similar host properties to unobscured TDEs ($\approx 11\%$) suggests that, as the authors posit, there is likely other variability at play that can drive MIRONG flares besides TDEs, including HV AGN-type flares. 

Finally, an additional estimate of the rate of TDEs in MIRONG can be made by again utilizing our finding from Section \ref{sec:dust} that the obscured hosts should mirror the unobscured population of TDE hosts. Briefly revisiting the bottom panel of Figure \ref{fig:fig1} and considering the subset of 5 \citetalias{2023ApJ...955L...6Y} TDEs that have matches in our host galaxy catalog, we see that very few of the MIRONG occupy the post-starburst region favored by TDEs and shown by the dotted black lines. Using a simple analysis, if 2 out of 5 TDE hosts (40\%) lie in this region, and 2 out of 103 MIRONG (1.9\%) occupy this region as well, we can estimate the number of obscured TDEs responsible for MIRONG behavior to be somewhere around 5 (4.9\%).

\section{Summary and Conclusions}

The arguments leading to the conclusion that the bulk of MIRONG are obscured HV AGN can be broadly summarized as follows:   

\begin{itemize}
    \item MIRONG host galaxies appear more similar to AGN host galaxies in the SFR - $M_\ast$ plane than galaxies or TDE hosts (Figure \ref{fig:fig1}). 
    \item Nuclear dust content does not appear to correlate with host galaxy type, stellar properties, or galaxy orientation. We also find no strong correlation between nuclear dust content with galactic-scale dust content as estimated by the Balmer decrement in MIRONG host galaxies showing no AGN activity. This suggests that the obscured host galaxy transient population should trace the unobscured one and that we are not missing a population of TDEs in highly SF galaxies (Figure \ref{fig:fig2}). This also implies that the observed post-starburst preference of TDEs is intrinsic, as opposed to an observational bias, raising many questions about the uniqueness of these systems. What is more, we conclude that the intrinsic rate of TDEs in star-forming galaxies is significantly lower than in post-starburst hosts.
    \item Based on a comparison with the unobscured population, we estimate the relative fraction of TDEs responsible for MIRONG to be $\lesssim 11\%$  (Figure \ref{fig:fig4}, see Appendix \ref{sec:appendix_b} for a summary of matching results).  
    \item  We conclude that the majority of MIRONG appear to be driven by HV AGN activity, as suggested by the higher  S\'ersic index distribution when compared to typical AGN (Figure \ref{fig:fig3}). 
\end{itemize}

\begin{figure}
\plotone{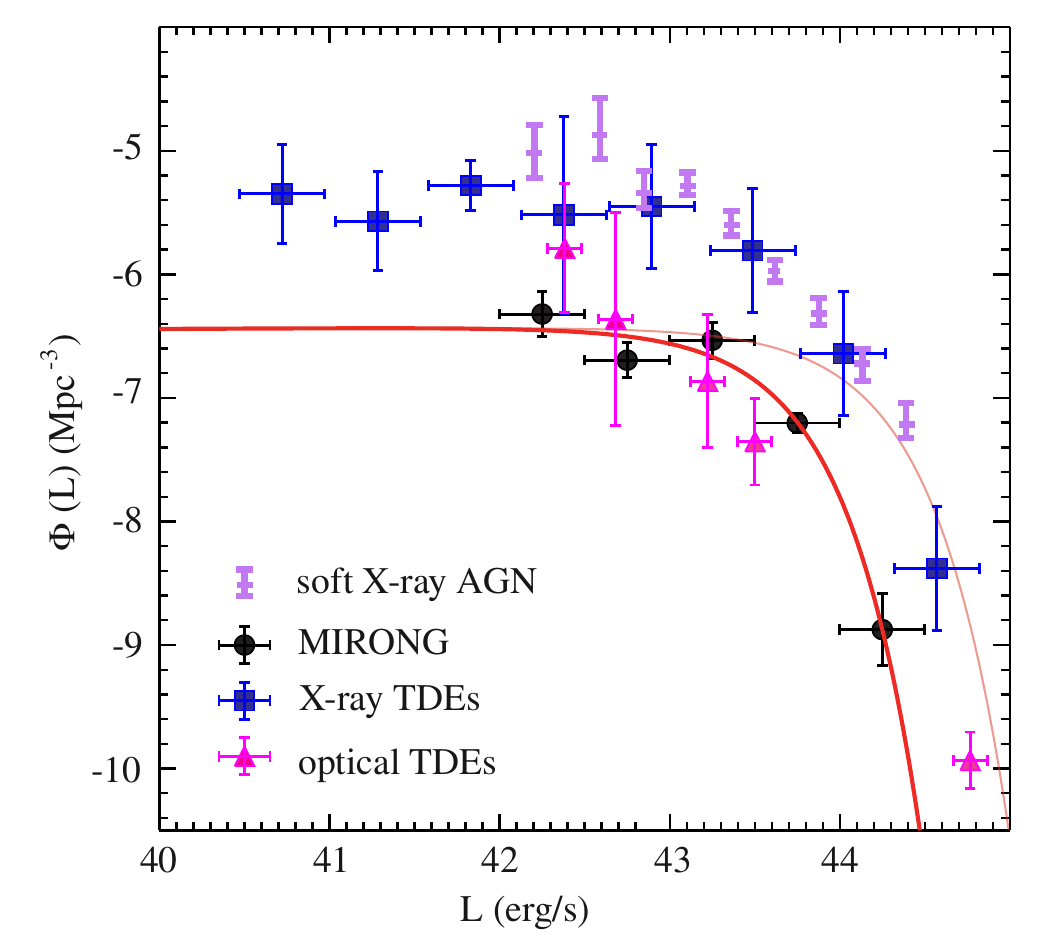}
\caption{Luminosity function for MIRONG for given luminosity bins, adapted from Figure 10 from \citet{2021ApJS..252...32J}. In addition to the X-ray TDEs from Figure 6 of \citet{2018ApJ...852...37A} and the optical TDEs from Figure 1 of \citet{2018ApJ...852...72V}, we additionally include the soft X-ray AGN sample from \citet{2018ApJ...852...37A}. The red line is a single Schechter function fit to MIRONG, and the fainter pink line represents the expected shift in the MIRONG distribution if a lower dust covering factor of 0.3 is assumed \citep[both from][]{2021ApJS..252...32J}.
\label{fig:fig5}}
\end{figure}

Having shown that TDEs are likely a smaller fraction of MIRONG than originally anticipated and that HV AGN might represent a higher fraction as shown by host galaxy and S\'ersic index analyses, we lastly consider whether or not the observed rates of these transient populations are consistent with one another. Figure \ref{fig:fig5} shows an adapted version of the luminosity function (LF) from \citet{2021ApJS..252...32J}, with the additional inclusion of a population of soft X-ray AGN from \citet{2018ApJ...852...37A}. The two lines represent fits to the MIRONG population for different assumed dust covering factors: the red line for a factor of 1, the fainter pink line for a factor of 0.3 \citep[both lines from][]{2021ApJS..252...32J}. As illustrated in \citet{2021ApJS..252...32J}, MIRONG appear similar in rate to optical and X-ray TDE fractions, especially given the uncertainty in the obscuration factor. However, we do note that other X-ray TDE rate estimates may be found to differ from the ones presented here \citep[see, e.g.,][]{2021MNRAS.508.3820S}. Soft X-ray AGN exhibit higher values of $\Phi$ (i.e. have a higher occurrence rate in the local universe) than MIRONG across luminosity bins. Given that HV AGN are by nature a smaller subset of the local AGN population, the LF of MIRONG would likely be consistent with that of HV AGN.  This is consistent with our findings that HV AGN could be responsible for most MIRONG.

\subsection{Relationship between nuclear obscuration and host galaxy properties}
The MIRONG population offers us an exciting opportunity to study the observational appearance of obscured  AGN and TDEs, which is not only determined by their intrinsic emission properties but also by the state of the intervening material along the line of sight. We find that nuclear dust obscuration does not correlate with host galaxy type or with the extinction
of the host galaxy. This is in agreement with \citet{2014MNRAS.437.3550M}, which finds no significant difference between the mean stellar masses and star formation rates of obscured and unobscured AGN hosts selected by X-ray flux. In essence, it is commonly understood that the physical state of intervening material is largely insensitive to the wider scale galactic conditions but appears to be mainly determined by radiation properties of the nuclear region \citep{2017Natur.549..488R,2020A&A...635A..92G,2021ApJ...922..179B,2022A&A...667A.140G}. For highly luminous AGN, the dust content inferred from the column of material responsible for the X-ray absorption is commonly larger than the one inferred from the reprocessing luminosity. This is consistent with the idea that X-ray absorbing gas is located within the dust sublimation radius, whereas the mid-IR flux arises from an area farther out \citep{2009ApJ...702.1127R}. Given that the MIRONG population, as argued in this {\it Letter}, is likely to be associated with highly luminous nuclear activity, we expect the X-ray absorbing region to be located within the dust sublimation radius. Swift J1200.8+0650 provides us with the best example to test this hypothesis, given that it was identified independently by the  MIRONG  and the high Galactic latitude {\it Swift} survey. The  X-ray absorbing gas measurement in Swift J1200.8+0650 \citep[$6-8\times 10^{22}$ cm$^{-2}$;][]{2007ApJ...669..109L} implies a much higher dust mass than the one derived from the mid-IR luminosity under the assumption of a high covering factor \citep[$\approx 0.2M_\odot$;][]{2021ApJS..252...32J}. As expected, the dust mass inferences in the MIRONG population are thus likely to be highly sensitive to the physics of dust sublimation. Be that as it may,  radiative feedback in Swift J1200.8+0650 and other MIRONG sources is still unable to completely clear out the circumnuclear dust environment. In sources for which the removal of dusty gas might be ultimately efficient, the nuclear source may decline in luminosity, giving rise to unabsorbed sources at lower luminosities.

In closing, here we confirm the general view that there is little physical connection between the gas accreting onto the SMBH and the material out of which stars form throughout the galaxy by demonstrating the lack of a connection between obscuring dust within the sphere of influence of the SMBH and the galaxy-wide properties (stellar age, stellar mass, SFR and dust content). This is consistent, for example, with the findings derived from large comprehensive studies using Herschel \citep{2012A&A...545A..45R} and XMM–Newton in the COSMOS field \citep{2014MNRAS.437.3550M}.

\acknowledgments
We thank the referee for incredibly useful comments and suggestions. We would like to express gratitude for insightful discussions with Julianne Dalcanton, Jenny Greene, Ryan Foley, and Morgan MacLeod. S.A.D. and E.R.R. thank the Heising-Simons Foundation, NSF (Graduate Research Fellowship, (AST-2150255 and AST-2307710)), Swift (80NSSC21K1409, 80NSSC19K1391) and Chandra (22-0142) for support. K.D.F. acknowledges support from NSF grant AST–2206164.

\appendix
\section{Host galaxy properties of MIRONG, CL AGN and AGN in the SDSS Catalog} \label{sec:appendix_a}

\begin{figure*}
\plotone{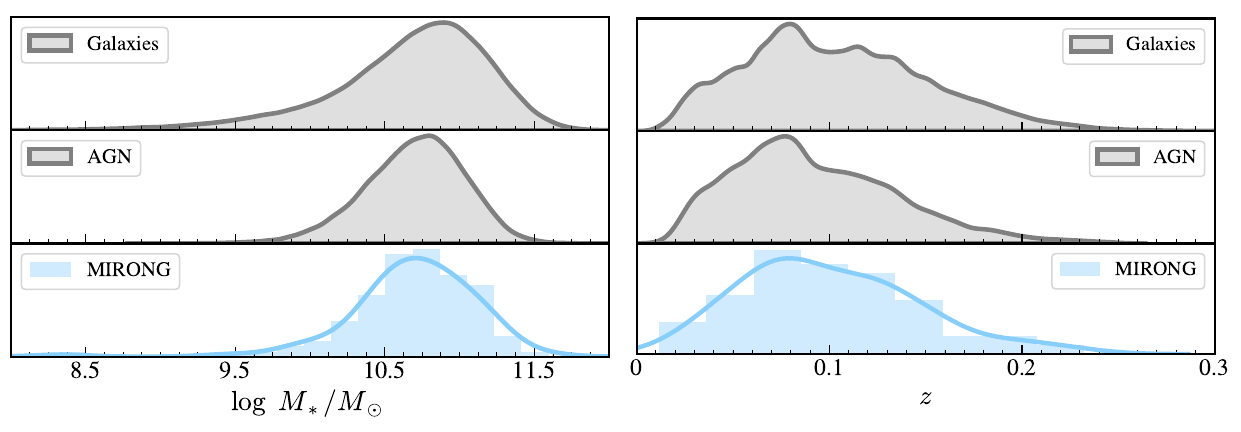}
\caption{Host galaxy comparisons in total stellar mass (left) and redshift (right) for all galaxies, AGN, and MIRONG in our catalog. The similarity between populations in each parameter makes possible the host galaxy comparison analysis that constitutes our methodology. 
\label{fig:fig6}}
\end{figure*}

The host galaxy matching analysis relies on the various methods used to uncovered the associated transient population. Since not all transients were uncovered using the same survey or discovery technique, an effective matching sample needs to be constructed. 
This is the case for our TDE sample, which we present in Section \ref{sec:tde_comp}. 
As mentioned briefly in Section \ref{sec:hosts}, this is not the case for MIRONG and CL AGN samples, which are primarily selected using the SDSS catalog.   Figure \ref{fig:fig6} gives credence to this idea and highlights the similarities in their distributions of total stellar mass (left) and redshift (right). We find that controlling for either of these properties before performing any matching or comparison described above does not yield differences in our conclusions, as expected given the similarities in their intrinsic distributions.

For completeness, we also refer the reader to Table \ref{tab:info} and and Table \ref{tab:tab1} for a summary of all MIRONG host galaxy properties referenced in this \textit{Letter} as well as for a synopsis of our CL AGN host galaxy matching exercise.

\section{TDE -- MIRONG Host Galaxy Matching Summary} \label{sec:appendix_b}
We perform two primary matching analyses in this work to constrain the fraction of MIRONG that may be TDEs: one in $^{0,0}u-r$ -- $\log\ M_\ast/M_\odot$ (yielding 10.8\%) and another in H$\alpha$ equivalent width (EW) -- Lick H$\delta_A$ absorption (yielding 4.9\%). The 10.8\% derived from the $^{0,0}u-r$, $\log\ M_\ast/M_\odot$ matching is the most robust given that it was performed with the largest samples of TDEs and MIRONG (21 and 37, respectively).

\begin{figure}
\epsscale{0.55}
\plotone{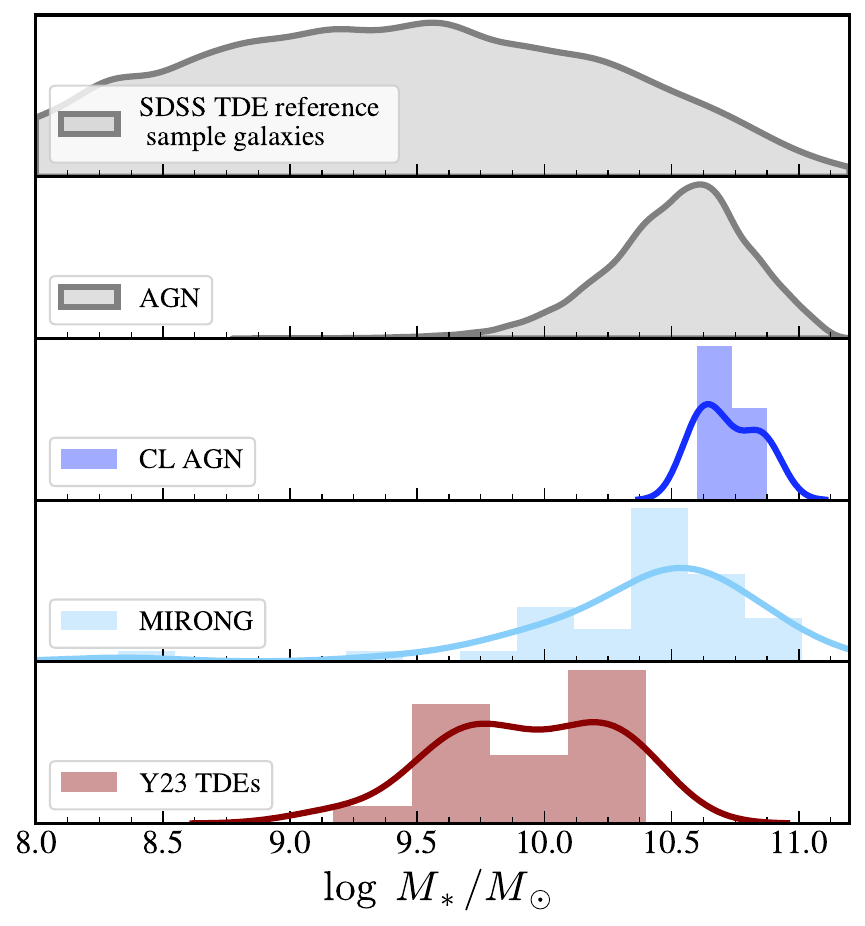}
\caption{Total stellar masses for all SDSS TDE reference samples (see Section \ref{sec:tde_comp} for a detailed explanation of sample selection), including galaxies, AGN, CL AGN, MIRONG, and \citetalias{2023ApJ...955L...6Y} TDEs. Even after controlling for selection effects unique to TDEs compared to the other transients, TDEs favor lower-mass host galaxies. 
\label{fig:fig7}}
\end{figure}

Figure \ref{fig:fig7} shows the distribution of total stellar masses for all SDSS TDE reference samples (see Section \ref{sec:tde_comp} for more detail). It is clear from this comparison that the TDEs favor lower-mass host galaxies. This continues to be highly prevalent even after we construct the matching sample in order to accurately account for detection biases. Matching MIRONG to TDE hosts in this reference sample yields 37.8\% taking only total stellar mass into account. This can be understood as an upper limit to the occurrence of TDEs in MIRONG hosts.

\startlongtable
\begin{deluxetable*}{ccccccccc}

\tablecaption{MIRONG Sample\label{tab:info}}
\tablehead{
\colhead{SDSS Name} & \colhead{R.A.} & \colhead{Dec} & \colhead{$z$} & \colhead{SFR} & \colhead{$\log\ M_\ast/M_\odot$} & \colhead{$\log\ M_{\rm dust}$} & \colhead{S\'ersic Index} & \colhead{$^{0,0}u-r$}  
}
\startdata
J110501.98+594103.5\tablenotemark{*a} & 166.2583	& 59.6843 & 0.0337 & 0.6938  & 10.316 & -1.58 & 5.15 $\pm$ 0.04 & 1.55 \\
J163246.84+441618.5\tablenotemark{*a} & 248.1952	& 44.2718 & 0.0579 & -1.3619 & 10.525 & -1.88 & 4.28 $\pm$ 0.17 &  2.52\\
J165726.81+234528.1\tablenotemark{*a} & 254.3617	& 23.7578 & 0.0591 & -0.3956 & 9.971  & -0.53 & 1.74 $\pm$ 0.11 & 1.89 \\
J104306.56+271602.1\tablenotemark{*a} & 160.7774	& 27.2673 & 0.1281 & 0.1142  & 10.654 & -1.25 & 7.96 $\pm$ 0.15 & 2.13 \\
J140221.26+392212.3\tablenotemark{*a} & 210.5886	& 39.3701 & 0.0638 & 0.3251  & 10.547 & -1.29 & 1.23 $\pm$ 0.03 & 1.92 \\
J151345.76+311125.0\tablenotemark{*a} & 228.4407	& 31.1903 & 0.0718 & 0.5689  & 10.829 & -1.55 & 2.50 $\pm$ 0.03 & 1.90\\
J164754.38+384342.0\tablenotemark{*a} & 251.9766	& 38.7283 & 0.0855 & 0.2108  & 10.214 & -1.82 & 1.85 $\pm$ 0.13 & 1.53\\
J154955.19+332752.0\tablenotemark{*a} & 237.4800	& 33.4644 & 0.0856 & 0.1408  & 10.017 & -1.36 & 1.27 $\pm$ 0.06 & 1.75 \\
J111536.57+054449.7\tablenotemark{*a} & 168.9024	& 5.7471  & 0.0900 & -0.8071 & 10.610 & -0.92 & 4.12 $\pm$ 0.22 & 2.40\\
J131509.34+072737.6\tablenotemark{*b} & 198.7889	& 7.4605  & 0.0918 & 0.7665  & 11.136 & -1.34 & 5.09 $\pm$ 0.10 & 1.96\\
J153711.29+581420.2\tablenotemark{*b} & 234.2971	& 58.2389 & 0.0936 & 0.3130  & 10.583 & -1.08 & 3.41 $\pm$ 0.19 & 2.14\\
J113355.93+670107.0\tablenotemark{*b} & 173.4831	& 67.0186 & 0.0397 & -0.4329 & 10.822 & -1.66 & 3.76 $\pm$ 0.02 & 2.32\\
J133212.62+203637.9\tablenotemark{*b} & 203.0526	& 20.6105 & 0.1125 & 1.0122  & 10.753 & -0.78 & 5.71 $\pm$ 0.18 & 1.79\\
J123852.87+081512.0\tablenotemark{*b} & 189.7203	& 8.2533  & 0.1138 & 0.4280  & 10.694 & -1.11 & 3.31 $\pm$ 0.14 & 1.86\\
J100350.97+020227.6\tablenotemark{*c} & 150.9624	& 2.0410  & 0.1247 & -0.0666 & 10.799 & -1.50 & 3.91 $\pm$ 0.23 & 2.24\\
J144227.57+555846.3\tablenotemark{*c} & 220.6149	& 55.9795 & 0.0769 & 0.8001  & 10.939 & -0.82 & 4.11 $\pm$ 0.02 & 2.02 \\
J124521.42 -014735.4\tablenotemark{*} & 191.3393 & -1.7932 & 0.2154 & 0.5375  & 11.050 & -0.58 & 7.81 $\pm$ 0.29 & 2.18 \\
110958.34+370809.6\tablenotemark{*} & 167.4931	& 37.1360 & 0.0260 & -1.1838 & 10.594 & -1.72 & 4.95 $\pm$ 0.01 & 2.63 \\
J155743.52+272753.0\tablenotemark{*} & 239.4314	& 27.4647 & 0.0316 & 0.9564  & 10.558 & -0.83 & 2.44 $\pm$ 0.01 & 2.51\\
112916.12+513123.5\tablenotemark{*} & 172.3172	& 51.5232 & 0.0329 & 0.3398  & 10.778 & -0.78 & 1.41 $\pm$ 0.01 & 2.44\\
J160052.26+461242.9\tablenotemark{*} & 240.2178	& 46.2119 & 0.1974 & 0.8476  & 11.191 & -0.88 & 1.64 $\pm$ 0.15 & 2.13\\
J010320.42+140149.8\tablenotemark{*} & 15.8351 	& 14.0305 & 0.0418 & 0.5445  & 10.929 & -0.81 & 1.01 $\pm$ 0.01 & 1.90\\
J083536.49+493542.7\tablenotemark{*} & 128.9020	& 49.5952 & 0.0424 & -0.3198 & 10.749 & -1.30 & 2.73 $\pm$ 0.02 & 2.34\\
J165922.65+204947.4\tablenotemark{*} & 254.8444	& 20.8298 & 0.0451 & 0.0911  & 11.144 & -1.46 & 2.97 $\pm$ 0.01 & 2.30\\
J120338.31+585911.8\tablenotemark{*} & 180.9097	& 58.9866 & 0.0469 & -0.5879 & 9.730  & -2.02 & 1.51 $\pm$ 0.13 & 1.62\\
J130815.57+042909.6\tablenotemark{*} & 197.0649	& 4.4860  & 0.0483 & -1.2743 & 10.627 & -1.85 & 4.78 $\pm$ 0.06 & 2.53\\
J105801.52+544437.0\tablenotemark{*} & 164.5064	& 54.7436 & 0.1306 & 0.6060  & 10.530 & -0.86 & 2.53 $\pm$ 0.24 & 1.57\\
J113901.27+613408.5\tablenotemark{*} & 174.7553	& 61.5691 & 0.1346 & 1.0686  & 10.605 & -0.84 & 0.95 $\pm$ 0.04 & 1.48\\
J162810.03+481047.7\tablenotemark{*} & 247.0419	& 48.1799 & 0.1245 & 0.7825  & 10.314 & -1.25 & 1.17 $\pm$ 0.06 & 1.56\\
J132259.94+330121.9\tablenotemark{*} & 200.7498	& 33.0227 & 0.1269 & 0.7953  & 10.893 & -1.10 & 1.03 $\pm$ 0.05 & 2.14\\
J112018.31+193345.8\tablenotemark{*} & 170.0763	& 19.5627 & 0.1278 & 1.1259  & 10.759 & -1.17 & 4.35 $\pm$ 0.22 & 1.63\\
J155437.26+525526.4\tablenotemark{*} & 238.6553	& 52.9240 & 0.0664 & -0.5625 & 10.502 & -1.27 & 5.97 $\pm$ 0.11 & 2.30\\
J155539.95+212005.7\tablenotemark{*} & 238.9165	& 21.3349 & 0.0709 & -0.7768 & 10.642 & -0.99 & 3.26 $\pm$ 0.12 & 2.59\\
J132902.05+234108.4\tablenotemark{*} & 202.2585	& 23.6857 & 0.0717 & -0.4885 & 10.565 & -0.94 & 1.97 $\pm$ 0.05 & 2.14\\
J130532.91+395337.9\tablenotemark{*} & 196.3871	& 39.8939 & 0.0725 & -1.2688 & 10.653 & -1.59 & 3.83 $\pm$ 0.12 & 2.41\\
J112446.21+045525.4\tablenotemark{*} & 171.1926	& 4.9237  & 0.0740 & 0.9605  & 10.729 & -0.83 & 2.08 $\pm$ 0.03 & 1.82\\
J094303.26+595809.3\tablenotemark{*} & 145.7636	& 59.9693 & 0.0749 & -1.2882 & 10.356 & -1.64 & 4.62 $\pm$ 0.28 & 2.43\\
J081403.78+261144.3\tablenotemark{*} & 123.5158	& 26.1956 & 0.0757 & 0.5499  & 10.917 & -0.86 & 2.67 $\pm$ 0.07 & 2.22\\
J150844.22+260249.1\tablenotemark{*} & 227.1843	& 26.0470 & 0.0826 & 0.3946  & 10.404 & -1.02 & 0.76 $\pm$ 0.03 & 1.86\\
J140648.43+062834.8\tablenotemark{*} & 211.7018	& 6.4763  & 0.0850 & 1.1048  & 10.770 & -1.04 & 1.32 $\pm$ 0.03 & 1.87\\
J141235.89+411458.5\tablenotemark{*} & 213.1496	& 41.2496 & 0.1025 & 0.0925  & 11.012 & -1.83 & 2.98 $\pm$ 0.12 & 2.48\\
J103753.68+391249.6\tablenotemark{*} & 159.4737	& 39.2138 & 0.1068 & 0.5869  & 10.446 & -0.98 & 1.10 $\pm$ 0.06 & 1.78\\
J090924.55+192004.8\tablenotemark{*} & 137.3523	& 19.3347 & 0.1072 & -0.2895 & 10.532 & -0.82 & 1.78 $\pm$ 0.10 & 2.26\\
J075709.69+190842.8 & 119.2904	& 19.1452 & 0.1050 & 0.6167  & 10.767 & -1.37 & 3.09 $\pm$ 0.12 & 2.06\\
J100120.37+182926.6 & 150.3349	& 18.4907 & 0.1060 & 0.0215  & 10.498 & -1.56 & 1.45 $\pm$ 0.06 & 2.15\\
J101708.94+122412.0 & 154.2873	& 12.4034 & 0.1076 & 0.0808  & 10.567 & -1.58 & 6.89 $\pm$ 0.33 & 1.83\\
J074547.87+265537.9 & 116.4495	& 26.9272 & 0.1148 & 0.9047  & 10.312 & -0.78 & 7.98 $\pm$ 0.11 & 1.53\\
J144024.32+175852.7 & 220.1013	& 17.9813 & 0.1157 & -0.1135 & 10.502 & -1.73 & 5.68 $\pm$ 0.39 & 2.17\\
J151257.19+280937.5 & 228.2383	& 28.1604 & 0.1155 & -0.8570 & 10.941 & -1.10 & 5.40 $\pm$ 0.27 & 2.52\\
J100809.02+154951.3 & 152.0376	& 15.8309 & 0.1176 & 0.3173  & 10.982 & -1.10 & 1.74 $\pm$ 0.06 & 2.02\\
J084752.78+514236.2 & 131.9699	& 51.7101 & 0.1200 & 0.8325  & 10.756 & -1.11 & 1.34 $\pm$ 0.10 & 1.75\\
J151117.94+221428.2 & 227.8248	& 22.2412 & 0.1205 & -0.8159 & 11.116 & -0.98 & 4.29 $\pm$ 0.30 & 2.42\\
J012100.67+140517.3 & 20.2528	& 14.0881 & 0.1294 & 0.8598  & 10.930 & -0.90 & 3.23 $\pm$ 0.18 & 1.85\\
J000046.46+143813.0 & 0.1936	& 14.6370 & 0.1366 & 0.5935  & 11.053 & -1.07 & 3.38 $\pm$ 0.18 & 2.45\\
J081451.87+533732.5 & 123.7161	& 53.6257 & 0.1390 & 0.8186  & 10.857 & -1.18 & 1.72 $\pm$ 0.10 & 1.94\\
J104138.79+341253.5 & 160.4117	& 34.2149 & 0.1403 & 0.7567  & 10.902 & -1.51 & 5.12 $\pm$ 0.28 & 1.93\\
J100955.70+220949.3 & 152.4821	& 22.1637 & 0.1415 & -0.5718 & 11.131 & -1.19 & 4.67 $\pm$ 0.19 & 2.33\\
J115326.76+403719.1 & 178.3615	& 40.6220 & 0.1451 & 0.8811  & 10.406 & -1.13 & 0.97 $\pm$ 0.04 & 1.54\\
J232452.26+154251.0 & 351.2178	& 15.7142 & 0.1511 & 0.6084  & 10.973 & -1.58 & 4.85 $\pm$ 0.38 & 1.85\\
J105344.12+552405.7 & 163.4339	& 55.4016 & 0.1517 & 1.1808  & 10.711 & -1.07 & 1.03 $\pm$ 0.05 & 1.62\\
J085959.47+092225.7 & 134.9978	& 9.3738  & 0.1519 & 1.1258  & 10.773 & -0.74 & 2.80 $\pm$ 0.10 & 1.76\\
J111431.83+405613.8 & 168.6327	& 40.9372 & 0.1525 & -0.7727 & 11.104 & -1.06 & 4.68 $\pm$ 0.43 & 2.51\\
J100256.90+442457.8 & 150.7371	& 44.4161 & 0.1545 & -0.5432 & 11.393 & -0.31 & 3.12 $\pm$ 0.11 & 2.72\\
J084157.98+052605.8 & 130.4916	& 5.4349  & 0.1563 & 0.3459  & 11.221 & -1.04 & 4.45 $\pm$ 0.24 & 2.12\\
J131022.77+251809.3 & 197.5949	& 25.3026 & 0.1604 & 1.1384  & 10.806 & -1.20 & 4.02 $\pm$ 0.35 & 2.16\\
J135241.36+000925.8 & 208.1724	& 0.1572  & 0.1660 & 0.9115  & 11.165 & -1.12 & 2.07 $\pm$ 0.09 & 2.08\\
J085434.65+111334.7 & 133.6444	& 11.2263 & 0.1672 & 0.8626  & 11.596 & -0.87 & 7.93 $\pm$ 0.07 & 1.97\\
J134105.98 -004902.5 & 205.2749 & -0.8174 & 0.1754 & 1.0146  & 11.074 & -1.11 & 6.15 $\pm$ 0.33 & 1.58\\
J120145.97+352522.5 & 180.4416	& 35.4229 & 0.1903 & 0.5322  & 11.192 & -0.68 & 5.32 $\pm$ 0.30 & 2.12\\
J112238.84+143348.4 & 170.6619	& 14.5634 & 0.1942 & 0.9404  & 11.252 & -1.06 & 7.96 $\pm$ 0.12 & 1.76\\
J104609.61+165511.4 & 161.5401	& 16.9199 & 0.2069 & 0.8473  & 11.346 & -0.81 & 8.00 $\pm$ 0.13 & 1.57\\
J102959.95+482937.9 & 157.4998	& 48.4939 & 0.2324 & 1.5759  & 10.655 & -0.75 & 3.90 $\pm$ 0.39 & 0.85\\
J154029.29+005437.2 & 235.1221	& 0.9104  & 0.0117 & -2.4216 & 8.322  & -3.71 & 1.09 $\pm$ 0.02 & 1.09\\
J154843.06+220812.6 & 237.1794	& 22.1368 & 0.0313 & -0.6606 & 9.945  & -1.42 & 3.80 $\pm$ 0.11 & 2.30\\
J120057.93+064823.1 & 180.2414	& 6.8064  & 0.0360 & -0.2476 & 11.058 & -0.70 & 3.38 $\pm$ 0.02 & 2.15\\
J142808.89 -023124.8 & 217.0371 & -2.5236 & 0.0521 & -0.7258 & 9.923  & -2.22 & 7.89 $\pm$ 0.38 & 2.04\\
J215648.45+004110.6 & 329.2019	& 0.6863  & 0.0539 & -0.3556 & 10.549 & -1.80 & 5.83 $\pm$ 0.20 & 2.28\\
J004500.47 -004723.1 & 11.2520  & -0.7897 & 0.0568 & -0.0044 & 9.426  & -2.20 & 1.19 $\pm$ 0.08 & 1.25\\
J150440.39+010735.0 & 226.1683	& 1.1264  & 0.1283 & 0.6755  & 11.187 & -0.64 & 3.79 $\pm$ 0.15 & 2.14\\
J095754.76+020711.2 & 149.4782	& 2.1198  & 0.1253 & 0.5030  & 10.736 & -1.42 & 3.62 $\pm$ 0.23 & 2.27\\
J134123.20+151650.4 & 205.3467	& 15.2807 & 0.1255 & 1.0196  & 10.906 & -1.03 & 1.99 $\pm$ 0.07 & 1.89\\
J120942.22+320258.8 & 182.4259	& 32.0497 & 0.0590 & 0.2593  & 10.784 & -1.36 & 4.44 $\pm$ 0.05 & 2.02\\
J140950.27+105740.2 & 212.4595	& 10.9612 & 0.0597 & 0.3989  & 11.246 & -0.83 & 1.82 $\pm$ 0.01 & 2.32\\
J114922.02+544151.4 & 177.3418	& 54.6976 & 0.0619 & 0.4912  & 10.566 & -1.72 & 1.17 $\pm$ 0.02 & 1.83\\
J084232.87+235719.6 & 130.6370	& 23.9555 & 0.0635 & 0.4795  & 10.464 & -2.08 & 4.99 $\pm$ 0.08 & 1.69\\
J105145.47+210132.1 & 162.9395	& 21.0256 & 0.0659 & -0.6158 & 10.304 & -2.02 & 4.54 $\pm$ 0.25 & 2.14\\
J144829.01+113732.1 & 222.1209	& 11.6256 & 0.0666 & 0.3213  & 10.711 & -1.36 & 2.28 $\pm$ 0.04 & 2.07\\
J142420.78+624916.5 & 216.0866	& 62.8213 & 0.1091 & 0.4550  & 10.960 & -1.56 & 3.58 $\pm$ 0.13 & 2.24\\
J081121.40+405451.8 & 122.8392	& 40.9144 & 0.0670 & 0.8102  & 10.217 & -1.17 & 5.98 $\pm$ 0.20 & 1.90\\
J153310.02+272920.2 & 233.2918	& 27.4890 & 0.0719 & -0.4336 & 10.972 & -1.11 & 4.96 $\pm$ 0.05 & 2.26\\
J161258.17+141617.5 & 243.2424	& 14.2715 & 0.0720 & -0.0619 & 10.510 & -1.39 & 6.26 $\pm$ 0.28 & 1.95\\
J020552.15+000411.7 & 31.4673  & 0.0699  & 0.0765 & 0.8150  & 10.345 & -0.94 & 1.42 $\pm$ 0.06 & 1.70\\
J143016.05+230344.4 & 217.5669	& 23.0623 & 0.0810 & -0.2164 & 11.228 & -1.33 & 3.88 $\pm$ 0.06 & 2.34\\
J121907.89+051645.7 & 184.7829	& 5.2794  & 0.0825 & 0.0719  & 10.226 & -1.31 & 2.19 $\pm$ 0.11 & 2.14\\
J152438.13+531458.7 & 231.1589	& 53.2496 & 0.0851 & 0.1311  & 10.011 & -2.02 & 1.36 $\pm$ 0.07 & 1.71\\
J214142.90 -085702.3 & 325.4288 & -8.9507 & 0.0873 & -0.3762 & 10.499 & -1.56 & 2.65 $\pm$ 0.15 & 2.23\\
J085835.90+412113.8 & 134.6496	& 41.3539 & 0.0870 & -1.0266 & 10.498 & -1.06 & 3.81 $\pm$ 0.26 & 2.44\\
J093135.46+662652.2 & 142.8978	& 66.4478 & 0.0873 & -1.0901 & 10.784 & -1.90 & 6.06 $\pm$ 0.28 & 2.23\\
J124255.36+253727.9 & 190.7307	& 25.6244 & 0.0879 & -1.3684 & 10.426 & -1.57 & 4.51 $\pm$ 0.25 & 2.51\\
J134032.49+184218.6 & 205.1354	& 18.7052 & 0.0902 & -0.5332 & 10.609 & -1.77 & 5.79 $\pm$ 0.24 & 2.14\\
J132848.45+275227.8 & 202.2019	& 27.8744 & 0.0911 & 0.1618  & 10.828 & -1.35 & 5.86 $\pm$ 0.21 & 2.00\\
J083721.86+414342.0 & 129.3411	& 41.7283 & 0.0981 & 0.6995  & 11.010 & -1.10 & 3.51 $\pm$ 0.12 & 2.20\\
J091531.04+481407.7 & 138.8794	& 48.2355 & 0.1005 & -0.7665 & 10.874 & -1.54 & 6.61 $\pm$ 0.12 & 2.42\\
\enddata
\tablenotetext{*}{Have multi-epoch spectroscopy from \citet{2022ApJS..258...21W}.}
\tablenotetext{a}{Tentative classification from \citet{2022ApJS..258...21W} as TDE.}
\tablenotetext{b}{Tentative classification from \citet{2022ApJS..258...21W} as AGN flare.}
\tablenotetext{c}{Tentative classification from \citet{2022ApJS..258...21W} as turn-on CL AGN.}
\end{deluxetable*}

\begin{deluxetable*}{ c c c c  }
\tablecolumns{4} 
 \tablecaption{CL AGN host galaxy matching summary. S\'ersic index errors are the median of individual S\'ersic errors. \label{tab:tab1}}
 \tablehead{
 \colhead{Category} & \colhead{Fraction} & \colhead{Percent} & \colhead{Median S\'ersic Index}  
}
\startdata
 All MIRONG & 18/103 & 17.5 & 3.96 $\pm$ 0.14 \\
 AGN MIRONG & 10/52 &  19.2 & 4.32 $\pm$ 0.14 \\
 AGN & 9,455/52,613 & 18.0 & 3.05 $\pm$ 0.08 \\
 Galaxies & 73,722/500,707 & 14.7 & 4.01 $\pm$ 0.15 \\
 \enddata

\end{deluxetable*}

\end{document}